\documentclass[pra,aps,twocolumn,showpacs]{revtex4}
\usepackage{graphicx}
\usepackage{dcolumn}

\begin{document}
\def\be{\begin{equation}}
\def\ee{\end{equation}}
\def\bearr{\begin{eqnarray}}
\def\eearr{\end{eqnarray}}
\def\la{\langle}
\def\ra{\rangle}
\def\l{\left}
\def\r{\right}

\title{Sound velocity and multibranch Bogoliubov spectrum of an 
elongated Fermi superfluid in the BEC-BCS crossover}

\author{Tarun Kanti Ghosh and Kazushige Machida}
\affiliation
{Department of Physics, Okayama University, Okayama 700-8530, Japan}

\date{\today}

\begin{abstract}
We study properties of excited states of an elongated Fermi superfluid
along the BEC-BCS crossover including the unitarity limit. 
Analytic expressions for the sound velocity in an inhomogeneous as
well as homogeneous Fermi superfluid along the crossover are obtained
on the basis of the hydrodynamic theory.
The complete excitation spectrum of axial quasiparticles with various 
discrete radial nodes are presented. We discuss the feasibility of 
measuring the sound velocity and the multibranch Bogoliubov spectrum 
experimentally.
\end{abstract}

\maketitle

\section{Introduction}

Strongly interacting two-component Fermi gases provide a unique
testing ground for the theories of exotic systems in nature.
In atomic Fermi gases, tunable strong interactions are produced
using the Feshbach resonance \cite{houb,stwa,ties}. 
By sweeping the magnetic field in the Feshbach resonance experiments,
magnitude and nature of the two-body interaction strength changes 
from repulsive to attractive.
Across the resonance the $s$-wave scattering length $a$ goes from large
positive to large negative values.
The fermionic system becomes molecular Bose-Einstein condensates (BEC)
for strong repulsive interaction and
transforms into the Bardeen-Cooper-Schrieffer (BCS) superfluid when the 
interaction is attractive.
The first observations of BEC of molecules consisting of loosely bound 
fermionic atoms \cite{greiner,jochim, zw} initiated a series of
explorations \cite{hara,regal,barten,bourdel,chin} of the crossover between 
BEC and BCS superfluid. 
The size of fermion pair condensates smoothly increases from the BEC to 
the BCS-side of the resonance.
Near the resonance, the zero energy $s$-wave scattering length $a$
exceeds the interparticle spacing and the interparticle interactions
are unitarity limited and universal.
Recent experiments have entered the crossover regime and yielded results 
of the interaction strength by measuring the cloud size and expansion.

As in the case of bosonic clouds, the frequencies of collective
modes of Fermi gases can be measured to high accuracy, it is
of major interest to investigate their dependence on the equation
of state along the crossover.
It was pointed out \cite{stringari} the 
collective frequencies of a superfluid Fermi gas at $T=0 $, 
trapped in a harmonic potential, approach well defined values
in the BEC and the unitarity limit regimes, where the density
dependence of the chemical potential can be inferred from general
arguments. In the intermediate region, various investigations, based
on the hydrodynamic theory of superfluid and suitable parameterizations of
the equation of state, have appeared recently 
\cite{hui,heiselberg,bulgac,kim,manini,combescot,astra1}. The first
experimental results on the collective frequencies of the lowest axial
and radial breathing modes on ultra cold gases of ${}^6$Li across the Feshbach
resonance have also become available \cite{kinast,bar}.
Since the BCS and the unitarity limits are characterized by the
same collective excitation frequencies, there is a growing
interest to study the sound velocity \cite{ho,heisel,tosi} to make a clear 
identification of these two regimes and to better characterize two kinds of
superfluid.

The axial excitations of ultra cold gases in a cigar shaped
trap can be divided into two regimes:
i) long wavelength excitations where wavelength is equal or larger
than the axial size, ii) short wavelength excitations where
wavelength is much smaller than the axial size. In the former case,
the axial excitations are discrete and the lowest breathing mode frequency
has been measured \cite{kinast,bar}. In the later case, the axial excitations can be  
described by a continuous wave vector $k$. However, the
finite transverse size of the system also produces a discreteness
in the radial spectrum. The short wavelength axial phonons with different number
of discrete radial nodes give rise to the multi branch Bogoliubov
spectrum (MBS) \cite{zaremba}. 

The inhomogeneous density in the radial plane determines the 
curvature of the mode spectrum. The effect of the inhomogeneous density 
in the radial plane decreases (since the radial
size increases) as we go from the molecular BEC side
to the weak-coupling BCS side for fixed number of atoms and the trapping potential.
We would expect that the MBS will be different in the different regimes
and it can be used to distinguish different superfluid regimes along 
the BEC-BCS crossover.

It should be noted that the axial excited state is coupled with the  
discrete radial nodes within a given angular momentum symmetry.
For example, when we excite the system to study the
sound propagation along the symmetry axis, this perturbation inherently
excites all other low energy transverse modes having zero angular momentum.
Similarly, the above arguments are also applicable to other low energy 
mode spectrum,
{\em e. g.} spectrum for the breathing mode.
To determine the various mode spectrum, we must take into account that the 
incident of the mode coupling between the axial quasiparticle states and 
the transverse modes.

In this work, we calculate the sound velocity in an inhomogeneous as well
as homogeneous Fermi superfluid along the BEC-BCS crossover. 
We also study the low energy MBS of a cigar shaped superfluid Fermi gas along the 
BEC-BCS crossover by including the mode coupling. It is important to study
such spectrum in view of the current Bragg scattering experiment \cite{davidson}
on the MBS of an elongated cloud of weakly interacting BEC.

This paper is organized as follows. In Sec. II, we calculate the 
transverse eigenfrequencies and its corresponding eigenfunction of 
an elongated Fermi superfluid along the BEC-BCS crossover. 
In Sec. III, we discuss about the equation of state of the Fermi 
superfluid. The sound velocity, phonon mode and
monopole mode spectrum are presented in Sec. IV. We give a brief
summary and conclusions in Sec. V.

\section{hydrodynamic equations and eigen-frequencies}
We consider a two-component Fermi gas in a long cigar shaped harmonic trap potential
$ V(r,z) = (M/2)(\omega_r^2 r^2 + \omega_z^2 z^2) $ at 
zero temperature. Here, $ \omega_z << \omega_r $. 
We assume that the system behaves hydrodynamically throughout all regime. 
If the system is BCS superfluid, then as long as  the oscillation frequency 
is below the gap frequency needed to break up a Cooper pair this condition 
is expected to be fulfilled. 
The system can be described by the following Schr$\ddot o$dinger equation \cite{kim}
\be
i \hbar \frac{\partial \psi}{\partial t} = 
[-\frac{\hbar^2}{2M} \nabla^2 + V(r) + \mu(n)] \psi,
\ee
where $M$ is the mass of the Fermi particles and $ \mu(n) $ is 
the equation of state which depends on the magnitude and
nature of the interaction strength. 

Using the Madelung transformation
$\psi = \sqrt{n} e^{i \theta} $ and neglecting the quantum pressure
term, we obtain the
hydrodynamic equations of motion for the Fermi superfluid which are given by the
continuity and the Euler equations, respectively,

\be
\frac{\partial n}{\partial t} = - {\bf \nabla} \cdot [n {\bf v}],
\ee
and 
\be
M \frac{\partial {\bf v}}{\partial t} = 
- \nabla[ \mu(n) + V(r) + \frac{1}{2} M {\bf v}^2].
\ee
Here, $ n({\bf r},t) $ and 
$ {\bf v}({\bf r},t) = (\hbar/M) \nabla \theta $ are the local 
density and superfluid velocity, respectively. 
We also assumed that 
$ \omega_r >> \omega_z $ so that it makes a long cigar shaped trap. 

The equation of state enters through the density-dependent 
chemical potential.
We assume the power-law form of the equation of state as 
$ \mu(n) = C n^{\gamma} $ as in Refs. \cite{hui,heiselberg,bulgac,manini,astra1}.
At equilibrium, the density profile takes the form
$ n_0(r) = (\mu/C)^{1/\gamma}( 1- \tilde r^2)^{1/\gamma} $, 
where $ \tilde r = r/R $ and
$ R = \sqrt{2 \mu/M \omega_r^2} $.
Linearizing around equilibrium, $ n = n_0 + \delta n $, $ {\bf v} = 
\delta {\bf v} $ and
$ \mu(n) = \mu(n_0) + (\partial \mu/\partial n)|_{n =n_0} \delta n $.
The equations of motion for the density and velocity fluctuations are
\be \label{den}
\frac{\partial \delta n}{\partial t} = - \nabla \cdot [n_0(r) \delta {\bf v}],
\ee
\be \label{vel}
M \frac{\partial \delta {\bf v}}{\partial t} = 
- \nabla [\frac{\partial \mu(n) }{\partial n}|_{n=n_0} \delta n].
\ee
Taking first-order time-derivative of Eq. (\ref{den}) and using Eq. (\ref{vel}),  
the second-order equation of motion for the density fluctuation is
given by
\be \label{den0}
\frac{\partial^2 \delta n}{\partial t^2} = 
\nabla \cdot [n_0(r) \nabla \frac{\partial \mu(n) }{\partial n}|_{n =n_0} 
\delta n ].
\ee

In the long cigar shaped trap, we assume the normal mode solution of 
the density fluctuation which can be written as
\be \label{plane}
\delta n(r,z,t) = \delta n(r) e^{i [\omega(k) t - k z]}.
\ee
Substituting Eq. (\ref{plane}) into Eq. (\ref{den0}), then one can obtain
\bearr \label{den1}
- \tilde \omega_{\alpha}^2(k) \delta n(r) & = & \frac{\gamma}{2} \nabla_{\tilde r} 
\cdot [(1- \tilde r^2)^{1/\gamma} 
\nabla_{\tilde r}(1- \tilde r^2)^{1-1/\gamma} \delta n(r)] \nonumber \\
& - &  \frac{\gamma}{2} \tilde k^2 (1-\tilde r^2) \delta n(r),
\eearr
where $ \tilde \omega = \omega/\omega_r $ and $ \tilde k = k R $. 
Here, $\alpha$ is a set of two quantum numbers: radial quantum number, 
$n_r$ and the angular quantum number, $m$.

For $ k = 0 $, it reduces to a two-dimensional eigenvalue problem and the solutions
of it can be obtained analytically. The energy spectrum is
given by
\be
\tilde \omega_{\alpha}^2 = |m| + 2 n_r [\gamma(n_r + |m|) + 1].
\ee
The corresponding normalized eigenfunction is given by
\be
\delta n_{\alpha} = A (1-\tilde r^2)^{1/\gamma -1} \tilde r^{|m|}
P_{n_r}^{(1/\gamma -1, |m|)} (2\tilde r^2 -1) e^{im\phi},
\ee
where $ P_{n}^{(a,b)}(x) $ is a Jacobi polynomial of order $n$ and $\phi $ is
the polar angle. Also, the normalization constant $ A $ is given by
\be
A^2 = \frac{2^{2-2/\gamma}}{\sqrt{\pi} R^2}
\frac{[\Gamma(n_r+1)]^2 \Gamma(1/\gamma)\Gamma(2/\gamma + 2 n_r + |m|)}
{\Gamma(1/\gamma-1/2)[\Gamma(1/\gamma + n_r)]^2 \Gamma(2n_r + |m| +1)}.
\ee
For $ \gamma = 1 $, the above energy spectrum and its corresponding eigenfunction 
exactly matches with results of Ref. \cite{graham}.
Note that the modes with $n_r =0 $ and $m \neq 0 $ do not depend
on the equation of state. This is because the flow in these modes
are incompressible and the internal energy does not change
during the oscillation period.
The radial breathing mode is $ \omega_1 = \sqrt{2(\gamma + 1)} \omega_r $ 
which
exactly matches with the result of Ref. \cite{heiselberg}.
The experimental results of the radial breathing mode \cite{kinast,bar} 
is well described \cite{heiselberg} by this analytic spectrum.

The solution of Eq. (\ref{den1}) can be obtained for arbitrary value of $k$ by
numerical diagonalization.
For $ k \neq 0 $, we expand the density fluctuation as
\be
\delta n = \sum_{\alpha} b_{\alpha} \delta n_{\alpha} (r,\phi).
\ee

Substituting the above expansion into Eq. (\ref{den1}), we obtain,
\bearr \label{density2}
0 & = & [\tilde \omega_{\alpha}^2 - [|m| + 2 n_r ( \gamma (n_r +|m|) +1)]
\nonumber \\ & - &
\frac{\gamma}{2} \tilde k^2] b_{\alpha} +
\frac{\gamma}{2} \tilde k^2 \sum_{\alpha^{\prime}} M_{\alpha \alpha^{\prime}} 
b_{\alpha^{\prime}}.
\eearr
Here, the matrix element $ M_{\alpha \alpha^{\prime}} $ is given by
\bearr \label{matrix}
M_{\alpha \alpha^{\prime}} & = & A^2 \int d^2 \tilde r
(1-\tilde r^2)^{2\gamma_0} \tilde r^{2+|m| + |m^{\prime}|} e^{i(m - m^{\prime}) \phi}
\nonumber \\ & \times & P_{n_r^{\prime}}^{(\gamma_0,|m^{\prime}|)}(2 \tilde r^2-1)
P_{n_r}^{(\gamma_0,|m|)}(2 \tilde r^2-1),
\eearr
where $ \gamma_0 = 1/\gamma -1 $.
The above eigenvalue problem (Eq. (\ref{density2})) is block diagonal with no overlap
between the subspaces of different angular momentum, so
that the solutions to Eq.(\ref{density2}) can be obtained separately in
each angular momentum subspace. We can obtain all low energy
multibranch Bogoliubov spectrum on the both sides of the Feshbach resonance
including the unitarity limit from Eq. (\ref{density2})
which is our main result.
Equations (\ref{density2}) and (\ref{matrix}) show that the spectrum depends
on the average over the radial coordinate and the coupling between
the axial mode and transverse modes within a given angular momentum symmetry.
Particularly, the coupling is important for large values of $k $.

\section{equation of state}
To calculate the sound velocity and the MBS,
we need to know how the adiabatic index $ \gamma $ depends on the two-body 
interaction strength. 
At zero temperature, the energy per particle of a dilute Fermi system can be
written as 
\be
\epsilon = \frac{3}{5} E_F \epsilon (y),
\ee
where $ E_F = \hbar^2 k_F^2/2M $ is the free particle Fermi energy and
$ \epsilon (y) $ is a function of the interaction parameter $y = 1/k_F a $.
In the unitarity limit ($y \rightarrow 0^{\pm} $) one expects that the 
energy per particle is proportional to that of a noninteracting Fermi gas. 
The fixed-node diffusion Monte Carlo calculation of Astrakharchik {\em et al}.
\cite{astra} finds $ \epsilon (y \rightarrow 0) = 0.42 \pm 0.01 $.
An analogous calculation of Carlson {\em et al}. \cite{carlson} gave
$ \epsilon (y \rightarrow 0) = 0.44 \pm 0.01 $. 
The calculation of Astrakharchik {\em et al}.
\cite{astra} is quite complete and gives the behavior of the energy
of the system across the unitarity limit. On the
basis of the data of Carlson {\em et al}. \cite{carlson}, Bulgac and Bertsch
\cite{bulgac} proposed the following behavior of $ \epsilon(y)$ near the
unitarity limit:
\be
\epsilon (y) = \xi -  \zeta y - \frac{5}{3} y^2 + O (y^3),
\ee
where $ \xi \sim 0.44 $ and $ \zeta = 1 $ for both positive and  
negative values of $y$. However, the data of Ref. \cite{astra} 
gives a continuous but not differentiable behavior of 
$ \epsilon (y) $ near $ y=0$ and it suggest $\zeta = \zeta_- = 1 $ in
the BCS regime and $\zeta = \zeta_+ = 1/3 $ in the BEC regime.
On the basis of the data of Astrakharchik {\em et al}. 
\cite{astra}, Manini and Salasnich \cite{manini} proposed the 
following analytical fitting formula of $ \epsilon(y)$
for all regimes in the BEC-BCS crossover including the unitarity limit:
\be \label{fit}
\epsilon (y) = \alpha_1 - \alpha_2 
\tan^{-1} [\alpha_3 y \frac{\beta_1 + |y|}{\beta_2 + |y|}].
\ee 
This analytical expression is well fitted with the data of Ref. 
\cite{astra} for a wide range of $y$ on both sides of the resonance. 
We shall use Eq. (\ref{fit}) for further studies in this work.
Two-different sets of parameters are considered in Ref. \cite{manini}: one
set in the BCS regime ($y<0$) and an another set in the
BEC regime ($y>0$). In the BCS limit, the values of the parameters \cite{manini} 
are $ \alpha_1 = 0.42 $, $ \alpha_2 = 0.3692 $, $\alpha_3 = 1.044 $,
$\beta_1 = 1.4328 $ and $ \beta_2 = 0.5523 $. In the BEC limit,
the values of the parameters \cite{manini} are
$ \alpha_1 = 0.42 $, $ \alpha_2 = 0.2674 $, $\alpha_3 = 5.04 $,
$\beta_1 = 0.1126 $ and $ \beta_2 = 0.4552 $.
The advantage of a functional parameterization of $ \epsilon(y)$ is that
it allows straightforward analytical calculations of several physical
properties. The chemical potential $\mu $ is given by \cite{manini}
\be \label{chemical}
\mu = \epsilon(n) + n \frac{d\epsilon (n)}{d n} = 
E_F [\epsilon (y) - \frac{y}{5} \epsilon^{\prime}(y)],
\ee
where $ \epsilon^{\prime}(y) = \frac{\partial \epsilon (y) }{\partial y} $.
One can extract an effective adiabatic index $\gamma $ and its dependence
on the scattering length $a$ by defining the logarithmic derivative as 
\cite{manini}
\be \label{gamma}
\gamma \equiv \bar \gamma = \frac{n}{\mu}\frac{d\mu}{dn}
= \frac{\frac{2}{3} - \frac{2y}{5} \epsilon^{\prime}(y) +
\frac{y^2}{15} \epsilon^{\prime \prime}(y)}{\epsilon (y) - 
\frac{y}{5}\epsilon^{\prime}(y)}.
\ee

The radial size of the Fermi system in all the regimes of the 
BEC-BCS crossover can be obtained from the relation: 
$ R = \sqrt{2 \mu/M \omega_r^2} $. From Eq. ({\ref{chemical}), 
one can obtain the radial size which is given by
\be \label{radial}
R = r_0 \sqrt{\epsilon (y) - \frac{y}{5} \epsilon^{\prime}(y)},
\ee
where $ r_0 = a_{\rm av} (24N)^{1/6} $ is the radial size of the free Fermi gas
in a harmonic trap potential \cite{butts},  
$a_{\rm av} = \sqrt{\hbar/M \omega_{\rm av}} $ and
$ \omega_{\rm av} = (\omega_r^2 \omega_z)^{1/3}$ is the average 
oscillator frequency of the trap potential. 
In the weak-coupling BCS limit, the ground state energy per particle is
$ \epsilon_{\rm bcs}(n) = (3/5) E_F $ and the chemical potential is 
$ \mu_{\rm bcs} = E_F $. The corresponding radius is 
$ R_{\rm bcs} =  a_{\rm av} (24N)^{1/6} = r_0 $. 
In the unitarity limit, the ground state energy per particle is
$ \epsilon_{\rm uni}(n) = (3/5) E_F \xi $ and the chemical potential is
$ \mu_{\rm uni} = E_F \xi $. The corresponding radius is
$ R_{\rm uni} =  a_{\rm av} (24N \xi^3)^{1/6} = r_0 \sqrt{\xi} $. 

\section{sound velocity and multibranch Bogoliubov spectrum}
\subsection{Sound velocity}
Before presenting the exact numerical results, we make some approximation
for a quantitative discussion. If we neglect the couplings among  all other 
modes in the $m=0$ sector by setting
$ l^{\prime} = (n_r, 0) $ in Eqs. (\ref{density2}) and (\ref{matrix}), one
can easily get following spectrum:
\be \label{per}
\tilde \omega_{n_r}^2 = 2n_r (\gamma n_r + 1) + 
\frac{\gamma}{2}(1-M_{n_r,n_r}) \tilde k^2.
\ee
In the limit of long wavelength, the $ n_r = 0 $ mode is phonon-like 
with a sound velocity
\be \label{sin}
u_1 = \sqrt{\frac{(2- \gamma)\gamma}{2} \frac{\mu}{M}}.
\ee
For $ \gamma = 1 $, it exactly reproduces the weakly interacting 
BEC results \cite{zaremba}.
This sound velocity is different from the result obtained in
Ref. \cite{tosi}. The reason for the difference is given below.
In Ref. \cite{tosi}, the sound velocity is obtained by simply integrating 
Eq. (\ref{den1}) on radial coordinates. In this work, we are multiplying
by the complex conjugate of $ \delta n $ on both sides of Eq. (\ref{den1})
and then integrating it on radial coordinates. Since the density fluctuation
at the lowest energy state is a function of the radial coordinate, the two
average procedure gives two different result. Note that the 
correct average procedure is considered in our work.
For the homogeneous Fermi system, the sound velocity can be obtain from 
Eq. (\ref{per}) by neglecting the $ M_{n_r,n_r} $ and it is
given by 
\be \label{sho}
u_1 = \sqrt{\frac{\gamma \mu}{M}}.
\ee 
The sound velocity in the inhomogeneous system is smaller by a factor 
of $ \sqrt{1-\gamma /2 } $ with
respect to the sound velocity in a homogeneous Fermi systems. This is
due to the effect of the average over the radial variable which can be seen
from Eqs. (\ref{density2}) and (\ref{matrix}). 

Using Eqs. (\ref{chemical}), (\ref{gamma}) and (\ref{sin}), 
the sound velocity in the inhomogeneous Fermi superfluid  
along the BEC-BCS crossover including the unitarity limit is given by
\be \label{soundin}
u_1 = v_F \sqrt{\frac{[\frac{1}{3} - \frac{y}{5} \epsilon^{\prime}(y) +
\frac{y^2}{30} \epsilon^{\prime \prime}(y)][-\frac{1}{3} +  \epsilon (y) 
- \frac{y^2}{30} \epsilon^{\prime \prime}(y)]}{[\epsilon (y) - \frac{y}{5}
\epsilon^{\prime}(y)]}},
\ee
where $ v_F = \sqrt{ 2 E_F/M} $  is the Fermi velocity. 
Similarly, by using Eqs. (\ref{chemical}), (\ref{gamma}) and (\ref{sho}),
the sound velocity in the
homogeneous Fermi superfluid along the BEC-BCS crossover including the 
unitarity limit is given by
\be \label{soundho}
u_1 = v_F \sqrt{ [\frac{1}{3} \epsilon (y) - \frac{y}{5} \epsilon^{\prime}(y)
+ \frac{y^2}{30} \epsilon^{\prime \prime}(y)]}.
\ee
Equation (\ref{soundho}) exactly agrees with the result of Ref. \cite{manini}.

In the molecular BEC limit, the sound velocity in the inhomogeneous 
bosonic systems can be written as $ u_m = \sqrt{\mu_m/2M_m} $, where
$ \mu_m $ is the chemical potential of the molecular BEC
and $ M_m = 2 M $ is the mass of a molecule.
The chemical potential $ \mu_m $ can be written as
$ \mu_m = 4 \pi a_m \hbar^2 n_m/M_m $, where
$ n_m = k_F^3/6 \pi^2 $ is the molecular density and
$ k_F $ is the Fermi wave vector.
Here, $ a_m = 0.6 a $ is the two-body scattering length
between two bound molecules \cite{petrov}.
A simple expression for
the sound velocity in the molecular BEC limit can be written  as
\be \label{soundmol}
u_m = v_F \sqrt{\frac{0.6}{12\pi} \frac{1}{y}}.
\ee

Using equations (\ref{soundin}), (\ref{soundho}) and (\ref{soundmol}), 
we plot the sound velocity along the BEC-BCS crossover in Fig. 1.
\begin{figure}[ht]
\includegraphics[width=9.1cm]{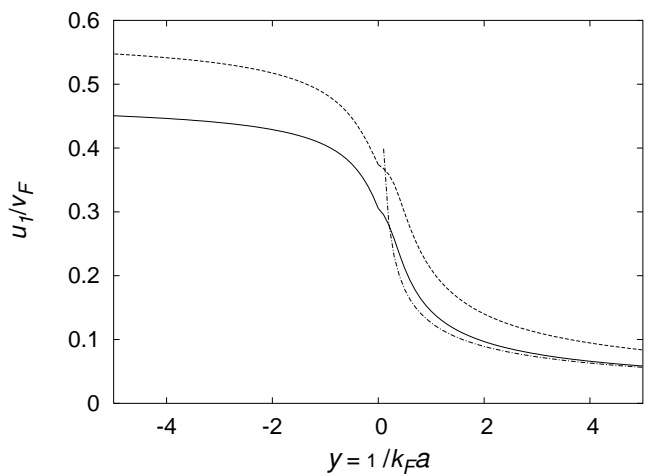}
\caption{Plots of the sound velocity along the BEC-BCS crossover including
the unitarity limit. The solid and dashed lines are corresponding to
the sound velocity in inhomogeneous
and homogeneous Fermi superfluid, respectively.
The dot-dashed line corresponds to Eq. (\ref{soundmol}).}
\end{figure}
There is a small kink at the unitarity limit 
$y=0$ due to $ \zeta_- \neq \zeta_+ $. Otherwise, Fig. 1 shows that 
there is a smooth crossover 
of the sound velocity from the molecular BEC side to the BCS side through 
the unitarity limit $ y = 0 $.  
Fig. 1 also shows that Eq. (\ref{soundmol}) matches very well with
Eq. (\ref{soundin}) for large values of $y$. 

For homogeneous Fermi systems the sound velocity in the two
limiting cases can be obtained from Eq. (\ref{soundho}) and 
these are given by $ u_1 = 0.37 v_F $ in the unitarity limit and by 
$ u_1 = 0.57 v_F $ in the weak-coupling BCS limit. These results 
exactly matches with the previous results \cite{ho,heisel}.
Similarly, the sound velocity for the inhomogeneous Fermi systems
in the two limiting cases can be obtain from Eq. (\ref{soundin})
and these are given by  $ u_1 = 0.30 v_F $ in the unitarity limit and 
$ u_1 = 0.45 v_F $ in the dilute BCS limit. 
The sound velocity in the inhomogeneous Fermi system
is less than that in the homogeneous Fermi system with the
same density at the center of the trap as the former system. However,
this difference is large in the BCS side compared to the BEC side.
The sound velocity of the inhomogeneous Fermi superfluid can be measured
by observing the propagation of the sound pulses along
the symmetry axis as it is done for weakly interacting BEC \cite{andrew}.  

\subsection{Phonon mode spectrum}
In Fig. 2 we plot the phonon mode spectrum  in the weak-coupling BCS limit 
($ y << 0 $), unitarity limit ($ y =  0 $) and BEC side of the
unitarity limit ($ y = 0.25 $) by solving Eq. (\ref{density2}).
These spectra have the usual form like $ \omega = u_1 k $  at low momenta, where
the sound velocity $ u_1 $ is given in Eq. (\ref{soundin}).
It is seen from Fig. 2 that the behavior of the phonon mode spectrum is different for 
different regimes characterizing each superfluid phase.
For example, the slope of the phonon spectrum in the BCS limit is large compared
to the unitarity and BEC limits as expected. 

\begin{figure}[ht]
\includegraphics[width=9.1cm]{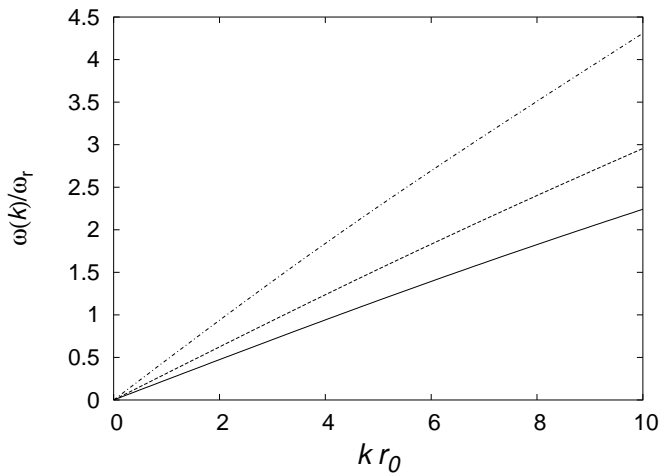}
\caption{Plots of the phonon mode spectrum in
the BCS-limit (dot-dashed), unitarity limit (dashed) and BEC side of the
unitarity limit with $ y = 0.25 $ (solid line).}
\end{figure}


\subsection{Monopole mode spectrum}
In Fig. 3, we plot the monopole mode spectrum in three different regimes
by solving Eq. (\ref{density2}).
In the long wavelength limit, the monopole mode has the free-particle 
dispersion relation with some effective mass $m_b $ and a gap 
$ \Delta_b = \sqrt{2 (\gamma + 1)} \omega_r $. In the long wavelength limit, 
the breathing mode spectrum can be calculated from Eq. (\ref{density2}) 
by using the first-order perturbation theory. The spectrum for the monopole
mode in the long wavelength limit is given by  
$ \omega_1(k) = \sqrt{2 (\gamma + 1)} \omega_r + \hbar k^2/2 m_b 
+ O (k^4)$,
where the effective mass of the breathing mode $ m_b $ is
\be
m_b = M \frac{\hbar \omega_r}{\mu} 
\sqrt{\frac{8}{\gamma^2} \frac{(2+\gamma)(\gamma+1)}{(2-\gamma)^2}}.
\ee
Note that $ \gamma = 2/3 $ in the BCS and unitarity limits.    
Therefore, the monopole mode frequencies are the same at the BCS and the
unitarity limits.  However, the  behavior of the spectrum in two 
different regimes are completely different. For example, the effective mass 
of the monopole mode spectrum in the
BCS limit is small compared to that of the unitarity limit.
\begin{figure}[ht]
\includegraphics[width=9.1cm]{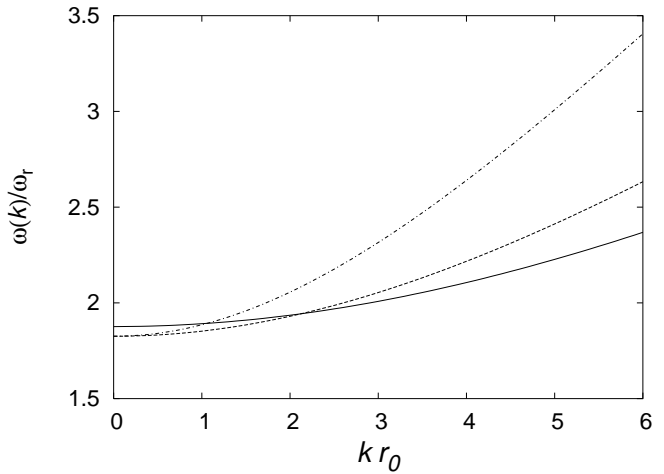}
\caption{Plots of the monopole mode spectrum in the BCS-limit (dot-dashed),
unitarity limit (dashed) and BEC side of the
unitarity limit with $ y = 0.25 $ (solid line).}
\end{figure}

\section{Summary and conclusions}
In this work, we have calculated the sound velocity in the
homogeneous as well as inhomogeneous Fermi superfluid along
the BEC-BCS crossover. 
The sound velocity in the inhomogeneous Fermi superfluid
can be measured by observing the sound pulse propagation
along the symmetry axis, similar to the experiment by 
Andrews {\em et al}. \cite{andrew} for weakly interacting BEC.
The hydrodynamic description presented in this
work enables us to produce correctly all low-energy multibranch
Bogoliubov spectrum by including the coupling of the axial mode
with the radial modes within the same angular momentum sector.
An analytic expression for the effective mass of the breathing 
mode spectrum is obtained. 
Due to the axial symmetry, the modes having zero angular momentum can 
be excited in the Bragg scattering experiment.
Particularly, the spectrum for the phonon and monopole modes in the different
regimes can be observed in the Bragg scattering experiments as
these spectrum are observed in Ref. \cite{davidson} for weakly interacting BEC. 
By measuring the sound velocity in the pulse propagation experiment and 
by observing the low energy Bogoliubov spectrum in the Bragg spectroscopy, 
one can make a clear identification of various superfluid regimes along 
the BEC-BCS crossover.

\begin{acknowledgments}
This work of TKG was supported by a grant (Grant No. P04311) of the
Japan Society for the Promotion of Science.
\end{acknowledgments}

\end{document}